# Leveraging policy instruments and financial incentives to reduce embodied carbon in energy retrofits

Haonan Zhang[1]

*School of Engineering, Faculty of Applied Science, The University of British Columbia, Kelowna, BC, Canada, V1V1V7*

**Abstract**

The existing buildings and building construction sectors together are responsible for over one-third of the total global energy consumption and nearly 40% of total greenhouse gas (GHG) emissions. GHG emissions from the building sector are made up of embodied emissions and operational emissions. Recognizing the importance of reducing energy use and emissions associated with the building sector, governments have introduced policies, standards, and design guidelines to improve building energy performance and reduce GHG emissions associated with operating buildings. However, policy initiatives that reduce embodied emissions of the existing building sector are lacking. This research aims to develop policy strategies to reduce embodied carbon emissions in retrofits. In order to achieve this goal, this research conducted a literature review and identification of policies and financial incentives in British Columbia (BC) for reducing overall GHG emissions from the existing building sector. Then, this research analyzed worldwide policies and incentives that reduce embodied carbon emissions in the existing building sector. After reviewing the two categories of retrofit policies, the author identified links and opportunities between existing BC strategies, tools, and incentives, and global embodied emission strategies. Finally, this research compiled key findings from all resources and provided policy recommendations for reducing embodied carbon emissions in retrofits in BC.

# 1 Introduction

Energy use is a main factor influencing the environmental sustainability of the building sector, and has received much attention with the ongoing climate action initiatives [2]. The existing buildings and building construction sectors together are responsible for over one-third of the total global energy consumption and nearly 40% of total direct and indirect greenhouse gas (GHG) emissions as per the International Energy Agency (IEA). The energy consumption and associated GHG emissions in the building sector may double and even triple by 2050. [3]. Poor energy performance of aging buildings in many countries makes a major contribution towards the said emissions. For example, according to the data from European Parliament, around 75% of existing buildings in Europe are energy inefficient [2]. In Canada, over 50% of Canadian residential buildings aged more than 30 years, and over 20% aged 50 years or more [4]. Old buildings mainly use non-renewable energy options and undergo material and component deterioration [5–7]. Thus, this older building stock consumes more energy and consequently emits more GHGs as compared to new construction [8,9]. Moreover, most of aging buildings do not comply with the latest energy-efficiency requirements [10,11].

However, the energy-efficient retrofit rates still remain low across the world. For example, retrofit rate is at an average of 1% in Germany [12], 1.0–3.0% in the UK [13], and 0.4-1.2% in Europe per annum [14]. In order to meet the target of emission reductions and the requirements of the latest energy-efficiency codes, immediate actions need to be taken to reduce the environmental impacts of the aging building stock [15]. As such, retrofits that involve modifications to envelope components and energy systems of existing buildings have garnered more attention in many countries [16].

With buildings becoming more energy-efficient, embodied carbon emissions play an increasing essential impact on achieving the goal of carbon reductions and climate changes. Embodied emissions can even become the primary source of carbon emissions in the building sector [17]. Implementing energy retrofits that can most reduce operational carbon emissions may not produce the benefit of life cycle carbon reductions because of higher embodied carbon emissions. According to the literature, the average share of embodied GHG emissions from buildings following current energy performance regulations is approximately 20–25% of life cycle GHG emissions. This figure can increase to 45–50% for highly energy-efficient buildings and surpasses 90% in extreme cases [1]. As such, it is important to mitigate the influence of embodied emissions when implementing energy retrofits.

Government-administered policy instruments are necessary to support the reduction of energy consumption and GHG emissions in the building sector [18]. In reality, retrofitting is an activity that involves multiple stakeholders, making it necessary to streamline and coordinate this process via regulation and policies [19,20]. Renovators and homeowners are mainly responsible for implementing retrofit measures for residential buildings, and there is a possibility of obtaining positive outcomes through retrofits. However, the various economic, environmental, and social issues associated with the entire process have to be addressed by the local or federal governments [21]. Homeowners might not embark on retrofitting without external economic support, especially when they plan to lease or sell their homes [22,23]. Furthermore, selecting the best retrofit options for a given context can vary significantly due to many factors, such as retrofit objectives, building types, climate contexts, technical issues, and stakeholder interests [24]. The benefits of retrofit policies in addressing the above issues are highlighted by literature [25–28]. For example, building assessment and certification policies can assist householders in understanding the energy performance of their homes and the retrofit benefits to their homes. In addition, financial incentives can reduce economic burdens on homeowners and encourage them to start retrofits [29–32].

This research aims to leverage existing policies to reduce embodied carbon in retrofits in BC. Firstly, this research conducted a literature review and identification of BC policies and incentives for reducing overall GHG emissions from the existing building sector. Then, this research analyzed worldwide policies and incentives that reduce embodied carbon emissions. After reviewing the two categories of retrofit policies, the authors identified links and opportunities between existing BC strategies, tools, and incentives, and global embodied emission strategies. Finally, this research compiled key findings from all resources and provided policy recommendations for reducing embodied carbon emissions in retrofits in BC.

## 2 Classification framework for retrofit policies

According to [33], policy instruments focus on stakeholders' main concerns and can be important nexus between policy objectives and application market. These instruments have been recognized as primary channels for administrations to develop social and economic guidelines. Retrofit Policy Instruments (RPIs) represent a set of policy measures, such as requirements, ordinance, certifications, and financial supports, to achieve desired targets (e.g. increase of retrofit rate) set by governments [34]. It has been proved that RPIs have great potential to be studied [21,28,35,36].

Therefore, to reflect the general characteristics of the RPIs, this paper presents a classification framework for the collected RPIs. By using this framework, the collected RPIs provided by governments, utility companies, and other relevant institutions are distributed into four categories: direction and command (DC), assessment and disclosure (AD), research and service (RS), and financial incentives (FI). The descriptions of the DC, AD, RS, and FI instruments are shown in Table 1.

**Table 1 Clusters of retrofit policy instruments**

| Instruments | Descriptions | References |
|---|---|---|
| DC | Overall direction: overall strategy, action plan; retrofit directives: environmental requirements, standards, regulations, and retrofit guidelines for existing residential buildings. | [26,27,37–39] |
| AD | Building energy performance assessment: energy auditing, rating, labeling, benchmarking, post-retrofit evaluation; disclosure of building energy consumption. | [26,40–43] |
| RS | Innovative researches: well-designed retrofit programs, new retrofit technologies, auxiliary tools; government public service: technical supports, increase of retrofit related institutions and departments, education and training programs. | [26,39,42–44] |
| FI | Economic supports for retrofit activities including grants: direct subsidies from governments; rebates: partial amount returned on applied retrofit measures; tax credits: deduction on the tax required to be paid; loans: purchase of retrofit materials or equipment at a low-interest rate. | [45–54] |

## 3 Analysis of retrofit policies

Municipalities have made great efforts to enforce a mix of different RPIs. These instruments can assist in improving stakeholders' retrofit awareness, knowledge, and skills and provide retrofit guidelines and supports for them [21,55].

### 3.1 Direction and command (DC) policies

When governments initiate RPIs, DC instruments, including direction and directive-based instruments, are their main choice, since the instruments have regarded as important interventions for retrofits at the initial stage [21]. The policies set overall retrofit strategies, targets, and requirements to assist stakeholders in better understanding basic knowledge, benefits, and the overall future direction for energy retrofits. Efforts have been made while considering stakeholders' interests and providing them with detailed retrofit guidelines. The efforts provide a clear framework for retrofit implementation, thereby overcoming retrofit implementation barriers and increasing market acceptance.

In response to climate change, the Clean BC Roadmap to 2030 has been proposed in British Columbia [56]. In the building sector, this plan includes the following important actions: achieving zero-carbon new construction by 2030, developing highest efficiency standards for new space and water heating equipment, enhancing energy efficiency programs, introducing home energy labelling, and using more low carbon building materials. This plan provides an overall direction for stakeholders to improve building energy efficiency and reduce GHG emissions in the building sector.

For existing buildings, the government of Canada has introduced three levels of retrofit measures, including minor retrofits, major retrofits, and deep retrofits [57]:

- Minor retrofits refer to modifications that are low-cost, easy to implement and that offer good value for the money and effort invested, such as upgrading lighting systems and adding insulation.
- Major retrofits generally take a more holistic approach to upgrade buildings, including replacing window glazing and doors, updating inefficient heating and cooling systems, installing low-flow faucets with sensors, automatic shut-offs, and sub-metering.
- Deep retrofits undertake an extensive overhaul of building's systems that can save up to 60 percent in energy costs, such as replacing the roof and replacing the heating, ventilation and air-conditioning system with a renewable technology like a ground-source heat pump. However, deep retrofits can be disruptive to building's occupants. Therefore, it is recommended to take the measures with tenant turnover or other major changes to occupancy.

DC instruments can play an important role in promoting retrofits by providing an overall development direction, retrofit requirements, and recommendations at the early stage of a retrofit project. With this kind of instrument, the stakeholders can be aware of long-term retrofit strategies, retrofit benefits, minimal retrofit requirements, and applicable retrofit measures. However, the compliance of DC instruments requires sufficient technological and financial support from well-trained retrofit practitioners, such as retrofit training courses, energy auditing data, and rebates for retrofit activities. In this regard, DC instruments should be combined with AD, RS, and FI instruments to improve the effectiveness of policy implementation.

## 3.2 Assessment and disclosure (AD) policies

AD instruments serve as an important tool for benchmarking building energy consumption, recognizing operational issues of systems, and identifying retrofit opportunities [58], [59]. This kind of instrument can help homeowners understand the operational energy use conditions and encourage energy retrofits. The related institutes provide energy audits for owners, including an examination of utility bills and comprehensive checks of targeted homes to identify energy-waste locations, and householders can solicit expert advice on retrofit options based on the auditing results. Furthermore, assessors can assist in ensuring that the energy saving goals are achieved after retrofitting.

The EnerGuide Rating System has been implemented across Canada to help Canadians improve the energy efficiency of their houses [60]. An EnerGuide rating is a standard measure of a home's energy performance. The home's energy efficiency level is rated on a consumption-based rating scale using gigajoules per year (GJ/year). The energy label also presents the breakdown of rated annual energy consumption, including space heating, space cooling, water heating, ventilation, lights and appliances, and other electrical energy consumption. Based on its energy efficiency, a home will receive an EnerGuide rating. Ratings are calculated by Registered Energy Advisors who analyze building plans, provide upgrade recommendations to improve energy efficiency, and complete a blower door test to confirm the air tightness of the home.

However, previous studies have indicated that the energy-efficiency labels might not be an essential consideration for landlords when they decide to rent or buy a property. Furthermore, the decisions are highly dependent on the disclosure of energy use data [61]. Therefore, disclosure policies have been introduced with the objectives of making occupants understand their energy use, arousing their environmental awareness, and encouraging them to upgrade the buildings with low energy efficiency through transparent information sharing [62]. Homeowners are required to report their energy uses periodically or when the buildings are put on the market [63]. However, it is recommended that multi-family and affordable buildings be excluded from the disclosure policies because of stakeholder perceptions that compliance would increase rent prices [64].

In addition, AD instruments can be used as effective tools to evaluate the differences between estimated and actual energy savings achieved from retrofits. Previous studies have proved that the discrepancy between expected energy use evaluated by energy modelling software and actual energy use is significant due to occupant behavior influences and inaccurate energy modelling, leading to unreliable paybacks for energy retrofit projects [65–67]. Developing simple occupant-based energy models that better address different occupant types and their impacts on energy use has been proven essential for improving modelling accuracy [68]. As such, AD instruments can help retrofit professionals understand why estimated and actual energy use are inconsistent and tune pre-retrofit energy modelling. After completing retrofits, energy savings can be recorded to test the accuracy of the developed energy model. However, the effective enforcement of AD instruments relies on financial supports and specialized retrofit professionals [28]. Therefore, FI and DS instruments should be developed to support the implementation of AD instruments. Furthermore, previous research has shown that it is not cost-effective to monitor energy data in a great level of detail [21]. Thus, striking a balance between building energy monitoring and modelling is essential.

### 3.3 Research and service (RS) policies

RS instruments, including research-related instruments and public service-related instruments, are used to increase occupants' awareness of energy retrofits and provided technological supports for retrofit practitioners.

Many efforts have been devoted to developing innovative research-related instruments, such as state-of-the-art retrofit technologies and innovative retrofit plans. The effectiveness of new technologies is mainly dependent on technological means, costs, climate conditions, and social contexts. In this regard, governments have launched research and development (R&D) programs to apply renewable energy systems to transfer the existing buildings to be a net-zero ready state.

Governments have provided many public services for stakeholders, such as establishing retrofit professional associations and developing training programs and retrofit support tools. Diverse institutions, departments, organizations, and workgroups, which can assist with coordinating among stakeholders and organizing retrofit activities, contribute to the adequate promotion, supervision, and service of home retrofit markets. A lack of practitioners with rich retrofit knowledge and skills is a big challenge that hinders the implementation of energy retrofits [69,70]. Training programs and support tools can strengthen practitioners' knowledge and skills associated with energy retrofits and provide convenience for them to initiate retrofitting and avoid troubles [24], [94]–[99]. Targeted training audiences generally involve occupants and professionals. Hargreaves et al. [77] suggested that occupants' energy-saving behaviors depend considerably on their inclination towards reducing energy consumption, information exchange, and skills to handle energy-efficient technical systems. The training programs for occupants are conductive to improve their awareness of upgrading their homes and increase their knowledge and skills on manipulating home appliances through an energy-efficient way [78]. In addition, retrofit practitioners' skills can also be enhanced through trainings. Previous practice has indicated that energy assessors trained in building science are more easily to gain homeowners' commitments to conduct retrofits than contractors trained in sales [37].

In terms of retrofit support tools, the EfficiencyBC program has been provided by the Province of British Columbia to assist occupants and business in accessing updated information about incentives and support to reduce energy use and GHG emissions in existing buildings [79]. The supports from this program include:

- Easy to use incentive search tools for residential and commercial renovations.
- Single application for EfficiencyBC, BC Hydro, FortisBC, and local government residential renovation incentives.
- Information and answers to frequently asked questions on energy efficiency upgrades.
- Free Energy Coaching Services for homeowners and businesses undertaking renovations, including a phone and email hotline staffed by energy coaching specialists.
- Search tool to find registered EnerGuide Rating System energy advisors for residential renovations.
- Contractor directories to find registered contractors in BC.

Clean BC Better Homes, which is an online rebate search tool, can assist homeowners in finding rebates according to their locations and building heating systems for residential buildings [80]. The heating systems consist of the following items: electricity (baseboard, furnace), natural gas or propane (furnace, boiler, fireplace), oil (furnace, boiler), wood (stove, fireplace), district energy (locally created energy), and heat pump. This program also provides a free coaching service for building owners and managers in BC. Energy Coaches are trained energy efficiency specialists who provide building-science based information about the options and opportunities to improve building energy efficiency. They are available to answer questions at all stages of energy retrofits project. The service, which can provide occupants with general advice about rebate programs and upgrade options, is described as follows:

- Access to Energy Coaches via a toll-free hotline and e-mail
- Information and general advice about energy efficiency upgrades and rebates
- Directing homeowners to appropriate program representatives

In addition, Clean BC Better Buildings, another online platform, can provide information and financial support for commercial buildings [81]. One of the custom programs under this sector is CleanBC Commercial Express Program, which guides and supports building operators and provides electrification opportunities across the commercial and institutional buildings [82]. They offer a free energy coaching service through fuel switching and other electrification measures. Program offers capital incentives based on age, location, square footage, hours of operation, and the type of equipment up to a maximum of $100,000 per project.

To conclude, RS instruments mainly rely on supports from local governments. These instruments also help stakeholders to explore new retrofit technologies and more easily access retrofit information and technical supports to address practical retrofit problems. In this sense, this type of policy can provide reliable supports for DC and AD instruments and disseminate the benefits brought by FI instruments, and thus, arouse stakeholders' awareness and improve their work efficiency.

### 3.4 Financial incentive (FI) policies

Financial incentives have been recognized as essential instruments for implementing retrofits in the residential buildings [83]. From the examinations of the practice of the surveyed policies in this paper, a multitude of financial incentives can be distributed into four groups: grants, rebates, loans, and tax credits.

In Canada, FI instruments are mainly provided by utility providers and local governments. A number of grant and rebate programs have been launched to encourage homeowners to upgrade building envelopes and energy systems.These programs can alleviate homeowners' concerns about high upfront costs of retrofits and reduce economic burdens on them [70,84]. For example, Natural Resources Canada (NRCan) provides grants for the combined cost of the pre-and post-retrofit evaluations up to a maximum of $600, which can assist occupants in evaluating retrofit outcomes. In addition, NRCan also provides direct grants for different retrofit measures as follows:

- Home insulation: Upgrade eligible attic, cathedral ceiling, flat roof, exterior wall, exposed floor, basement and crawl space (up to $5,000).
- Air-sealing: Perform air sealing to improve the building airtightness to achieve the air-change rate target (up to $1,000).
- Windows and doors: Replace doors, windows or sliding glass doors with ENERGY STAR® certified models (up to $5,000).
- Thermostat: Add a smart thermostat to help improve occupants' comfort and save money on energy bill (must be combined with another energy efficiency retrofit, up to $50).

- Space and water heating: Make the switch to more energy-efficient space heating or water heating equipment to save on utility bill and reduce carbon footprint (up to $5,000).
- Renewable energy: Install a solar photovoltaic system to convert sunlight energy into electricity (up to $5,000).
- Resiliency measures: Implement measures to protect home from environmental damages (must be combined with another energy efficiency retrofit, up to $2,625).

In British Columbia, Fortis BC provides rebates for purchasing energy-efficient HVAC equipment and appliances and replace old ones [85]. The product rebates include boilers, clothes dryers and washers, combination system, fireplaces, furnaces, heat pumps, natural gas heating system, refrigerators, thermostats, and water heaters. In the same vein, BC Hydro also offer rebates for energy-efficiency upgrades, including windows and doors (up to $3,000), insulation (up to $5,500), heat pumps (up to $2,000), and heat pump water heaters (up to $1,000) [86]. Furthermore, the Energy Efficiency Retrofit Program provided by BC Housing offers stakeholders additional funding to complete small-scale, energy saving retrofits of items such as light fixtures and boilers. Using all incentives together, stakeholders can undertake their retrofit projects and retain a portion of the ongoing energy savings [87].

Space heating has been recognized as the largest use of energy in homes. A properly installed heat pump is two to three times more efficient than other alternatives and can provide occupants with both comfortable heating in the winter and cooling for today's hot summers [88]. Thus, Clean BC Better Homes Low-interest Financing program has been introduced, which provides loans with interest rates as low as 0% for switching from a fossil fuel (oil, propane or natural gas) heating system to a heat pump [88]. Furthermore, the Canadian government has also implemented special programs to help low-income householders and senior citizens to live more comfortable, such as the "Healthy Homes Renovation Tax Credit" program and the "Energy Efficiency Retrofit Program for Low-Income Households" program.

As most of retrofit activities require economic supports, FI instruments have garnered much attention. These economic supports mainly come from government capitals, bank loans, utility provider rebates, and commercial institution grants. These instruments can address stakeholders' main concerns on economic issues related to retrofit projects, such as high initial cost, long payback period, and uncertainty of return on investment, and thus, improve their willingness and aspiration to upgrade buildings. In this regard, it is important to provide FI instruments to support the enforcement of DC, AD, and RS instruments. However, FI instruments might impose a heavy financial burden on governments, especially for local governments in a long run. Therefore, the level of investment depends on each governments' economic situation [89]. In addition, the confusing mix of FI instruments provided by different governments and utility companies as well as a lack of monitoring departments for financing retrofit projects also pose challenges to the effectiveness of FI instruments [90].

## 4 Analysis of global policies to reduce embodied emissions

The official energy efficiency departments in many countries have enforced a mix of different policies to reduce embodied emissions in the building sector. Different categories of policies are illustrated in the following sections.

### 4.1 Building regulations

1. Construction materials efficiency declaration

This policy requires declaration of key material mass per $m^2$/ square foot of building to be filed for the planning permit process at the occupancy permit application stage for buildings. As such, retrofit professionals can establish a benchmark database on building materials use and make stakeholders understand cash savings via saving material.

2. Expedited permitting for low carbon projects

This policy requires projects that meet given embodied carbon criteria should be given expedited or reduced fee processing. Examples of embodied carbon criteria may include life cycle assessment requirements or other requirements. For example, the City of Seattle's Priority Green Expedited program sets thresholds for energy efficiency, water conservation, waste reduction, and indoor air quality. Municipalities provide building owners who meet the requirements a single point of contact in the Department of Construction & Inspections with priority in scheduling an intake appointment, faster initial review of construction plans, and faster permit processing.

3. Prohibiting extremely high emitting materials

This policy prohibits the use of specific building materials associated with extremely high GHG emissions, such as spray foams with hydrofluorocarbon blowing agents used in insulation. Governments can implement different measures to ban the use of extremely high GHG-emitting building materials as follows:

- Ban the sale or purchase of the identified materials.
- Require designers or builders to specify defined low GHG-emitting alternative materials in order the permit to be approved.
- Financially penalize the use of high GHG-emitting materials and incentivize the use of low GHG-emitting alternative materials.

4. Life cycle carbon calculation and reporting.

This policy requires all projects to calculate and report their life cycle carbon emissions using a standardized measure, separating embodied and operational carbon.

### 4.2 Procurement policies

1. Carbon limits for building materials procurement

This policy set carbon intensity limits for key materials for construction materials used for retrofit projects and implement in public procurement. These can be demonstrated using Environmental Product Declarations (EPDs). For example, in Norway, the City of Trondheim requires the following from materials procurement:

- Concrete, both ready-mix and elements must meet low carbon class A
- Massive wood carbon limit 70 kg $CO_2/m^3$ - CLT and glulam carbon limit 100 kg $CO_2/m^3$
- Reinforcement steel shall be 100 % recycled, or emissions for other products shall be correspondingly lower
- Leveling screed on top of wooden floors must be B10, low carbon class A
- Non load bearing internal walls avg. emissions: max 10 kg $CO_2/m^3$

The data used to demonstrate compliance should be EPDs in compliance with EN 15804, and EPDs must be demonstrated for at least 15 products used in the building.

2. Requirement of recycled aggregates

This policy sets a minimum level of recycled or reused aggregates and soils in projects, if available within a predefined sourcing radius. For example, in Denmark, the government of Copenhagen requires in their Sustainability in Construction and Civil Works that "Road-building works must use crushed builders" rubble as a substitute for base gravel, provided that this is technically or economically sustainable. The crushed rubble must not contain any bricks, tiles or concrete that could be reused instead. In addition, France has established a target to achieve a 50 % share of reused or recycled building waste materials in road construction for materials bought by authorities in 2017, increasing to 60 % by 2020. This can help in having a market for most of the recycled aggregates processed from construction and demolition waste.

3. Require use of certified wood products

This policy requires the use of certified wood products when appropriate in projects for which municipal procurement guidelines apply. The required certification system should be determined by the governments and should have standards that have been demonstrated to produce wood with a lower embodied carbon footprint.

4. Circular materials purchasing strategy

This policy implements a strategy to define procurement in a manner which ensures that the market will either certainly or very likely deliver a circular solution in response. Procurement can be designed to focus on materials efficiency, circularity, maintainability, repairability and end of life opportunities.

### 4.3 Waste and circularity policies

1. Mandatory pre-demolition audits and data sharing

This policy requires all demolition and larger project permit applications to include a detailed pre-demolition audit. Municipalities make those pre-demolition audits public and allow a waiting time during which materials salvaging operators can recover what materials they commercially agree to recover from the building owner as opposed to instant demolition. For example, in Europe, the guidelines for the waste audits before demolition and retrofit works of buildings is a voluntary European protocol for pre-demolition audits of buildings that intend to help ensure recovery of recyclable material streams. It supports achieving high quality separated waste fractions.

2. Information on adaptability and waste reduction

This policy can provide designers, builders, and developers with information on financial benefits to be gained from reducing waste and information on converting existing buildings for adaptive reuse projects by selecting, procuring, and building with low-embodied carbon materials, and designing with later adaptation and reuse of buildings and materials. This can be accomplished through technical assistance, including educational workshops and trainings, through dissemination of financial analyses for low-embodied carbon project precedents, cost charts, as well as information on available tax incentives for donations of salvaged material or through the establishment of an office, internal resource center, or coordinator position to connect builders with information and expertise.

3. Materials longevity policy

This policy sets prescriptive requirements for the long-lasting design and use of long-lasting building materials. Further research must be conducted by the enforcing jurisdiction to determine the exact requirements, balancing the expected life span of a given material with its initial embodied carbon footprint, its potential for later reuse or recycling, and also the cost, availability, and local needs. For example, PVC windows have a significantly shorter lifetime than wood-aluminum windows. This policy should not apply to any buildings with planned short life-time.

## 4.4 Financial incentives

1. Provide tax rebates for low carbon development

Governments can offer an annual property tax rebate of up to 100% for a set number of years to property owners who opt for low-carbon renovation rather than new construction. The amount of the rebate can be based on a quantification of the embodied carbon reduction, so that projects with greater relative embodied carbon reductions are eligible for larger rebates.

2. Link land use fees to project life cycle carbon

Municipalities often charge land use fees from projects. These fees could be indexed to project embodied carbon. The charge structure could also be set up so that very low carbon projects would not pay fees at all or be possibly eligible for cash refunds.

Land use fees assessed for projects shall be established based on project life-cycle carbon intensity per square meter relative to building type benchmark. Project life-cycle carbon intensity per square meter is assessed using methodology and scope defined by the city. The life-cycle carbon intensity for project shall be verified by a competent verifier.

3. Provide Carbon performance grants for projects

Governments can set aside funds to award performance-conditioned grants for projects that achieve a clearly above market embodied carbon performance. Grants can be applied for during planning permission application, but they would be paid out only once project is completed, and performance achieved is possible to verify and audit.

4. Include embodied carbon in climate action plan

This program requires all future climate action plans or updates to existing climate action plans, to include an assessment of embodied carbon emissions from building and infrastructure construction, transportation, and land use. Also, to include a timeline and strategies for meeting reduction targets for embodied carbon in conjunction with timelines for reducing operational emissions.

5. Increase demolition permitting fees

Municipalities can increase demolition permitting fees for property owners applying to demolish buildings. Such increases could be applied conditionally, depending on building age, building size, predominant materials, carbon efficiency of the proposed replacement project, suitability of the building to deconstruction or other variables.

6. Create incentives for manufacturers to reduce carbon

This policy would create incentives for manufacturers located in the region of the jurisdiction to reduce embodied carbon in their products. Possible pathways include:

- Property/council tax rebates for manufacturers within the jurisdiction that demonstrate and quantify significant embodied carbon reductions in their main products.
- Property/council tax rebates for manufacturers within the jurisdiction who meet a Zero Net Carbon standard for the operation of their facilities, either by generating enough renewable, non-GHG emitting energy on-site to power their manufacturing processes, or procuring renewable, non-GHG emitting energy generated off-site.
- Direct grants/rebates for manufacturers for completing facility upgrades that significantly reduce carbon emissions, such as switching from a material or emissions intensive manufacturing process to a less material or emission-intensive alternative, or installing on-site renewable, non-GHG emitting energy generation.

Dedication of government resources and/or staff time to building relationships and facilitating the development of networks of manufacturers to support industrial symbiosis.

7. Establish landfill tax on construction and demolition waste

This policy establishes a requirement for taxing all landfilled construction and demolition waste. Landfill tax will provide a broad financial incentive to avoid final disposal of all types of material streams. To have impact on construction and demolition waste, this must also be levied on aggregates.

## 5 Recommendations for reducing embodied carbon in retrofits in British Columbia

After reviewing existing retrofit polices in British Columbia (BC) and polices that reduce embodied emissions across different countries, the following recommendations for building retrofit policies to reduce embodied emissions in British Columbia can be made.

### 5.1 Providing embodied carbon in retrofits related knowledge and information to the public

Knowledge and information associated with embodied carbon emissions play an essential role in the promotion of energy retrofits in the existing building sector. The knowledge on how retrofit technologies work can increase confidence of occupants in building energy retrofits. It is helpful to promote energy retrofits by providing information (e.g., potential benefits) to the public. From the perspective of energy retrofit investors, it is important for them to know the economic benefits of energy retrofits and what financial support is available. More importantly, such knowledge and information have a positive impact for enhancing occupants' willingness to performing energy retrofits. In British Columbia, as less knowledge and information are provided to the public, their awareness of building energy retrofits and building embodied emissions is at a low level, which lead to a low willingness to implementing energy retrofits. The situation poses a huge challenge to reduce embodied carbon emissions in retrofits in BC. In this regard, knowledge and information polices on building energy retrofits and building embodied emissions should be formulated in BC to increase public awareness.

In addition, it is essential for people to be able to obtain embodied emissions in retrofits related knowledge and information easily. For example, it is recommended that the public media (e.g., broadcast and TV) be used to disseminate knowledge and information of energy retrofits and building embodied emissions. Second, it is recommended that pilot retrofit projects be implemented. The carbon emission reduction data can be shared with the public so that the public can see the real benefits of building energy retrofits. Third, because of the fragmentation of relevant information, it is essential to establish a one-stop-shop information website where the public can easily obtain relevant knowledge and information at a low cost (in terms of effort and time).

| Advantages | Disadvantages |
|---|---|
| • It is a cost-effective way to improve occupants' awareness to implement energy retrofits for their homes.<br>• It can have an essential impact on reducing carbon emissions through retrofits. | • It can be time-consuming to implement this strategy.<br>• It cannot achieve direct carbon emission reductions.<br>• It may require special financial supports from governments or utility providers. |

## 5.2 Enhancement of building material performance

It is suggested that replace existing conventional construction materials with lower carbon and carbon sequestering materials. In addition, reducing overall material use by modifying the fabrication of individual building components and designing for material efficiency at the whole building scale is another effective way to reduce embodied carbon emissions. Building professionals can conduct analysis of building stock to identify low embodied carbon materials available in BC as alternatives to the current high embodied emission materials. (e.g., hempcrete and straw bale insulation can replace XPS foam). They can also conduct a cost-benefit analysis to estimate the potential of reducing embodied emissions in district buildings. Additional strategies for reducing embodied carbon include:

- Selecting natural products or those with low energy manufacturing processes. For example, timber or materials with natural fibers come from renewal sources and can be used with low processing. However, adding finishes to protect these materials increases their overall impact. Some varnishes for wood can limit its recyclability and lead to its use as energy source.
- Specifying durable materials suitable for the climatic context of BC. For example, the façade and roof are under constant wear from natural elements that can lead to frequent repairs and maintenance. By using durable materials, occupants can reduce the cost and frequency of refurbishment and the use of material replacement and its associated carbon footprint.
- Minimizing manufacturing and construction waste through comprehensive design (e.g., prefabricated panel manufacturing). The embodied carbon of a building element includes its material footprint and the waste that was generated during its construction. Prefabrication under controlled conditions allows reduction of waste and its associated carbon emissions. Similarly, modular elements permit the efficient use of materials and facilitate the industrialization/prefabrication of these elements.
- Selecting salvaged and recycled materials, materials that sequester carbon (e.g., using carbon capturing technology), and sourcing local supplies to avoid transportation emissions.
- Designing for end-of-life deconstruction and material reuse or recycling, as well as reducing material usage.

In terms of building material choice, insulation choice is among the most substantive opportunities for building retrofit planners to influence a building's life cycle emissions. Some insulating materials like straw bale, hempcrete, and wool store (sequester) carbon and have negative emissions, while others like extruded polystyrene (XPS) are made with blowing agents that have high global warming potentials (GWP). Compared with rigid insulation and spray foams, blown-in fibreglass and cellulose insulation have much lower carbon impacts. In addition, refrigerant choices also have material impacts on the carbon emissions associated with a building. HFC-134a is commonly used in air-conditioning systems; it has a GWP of 3,830 over a 20-year time horizon and GWP of 1,430 over 100 years. Some naturally occurring compounds that can be used as refrigerants, such as ammonia or propane, have much lower GWP. Refrigerant leakage rate is another important element that needs to be factored in when designing and selecting HVAC systems as it can have major impacts on a building's life cycle carbon emissions.

| Advantages | Disadvantages |
| --- | --- |
| <ul><li>The most direct and easiest way to reduce embodied carbon emissions in the existing building sector.</li><li>This strategy can produce life cycle cost benefits.</li></ul> | <ul><li>It may be a strategy with high upfront cost and long payback period.</li></ul> |

### 5.3 Engaging with energy retrofit professionals and organizations

As a lack of highly skilled retrofit professionals might undermine the effectiveness of retrofit schemes, increasing industry capacity is also essential. For example, recruiting newly qualified renovators by examining contractors who have work experience with old house retrofit or new energy-efficient dwelling construction should garner more attention. Practitioners could be encouraged to document their work qualifications and corroborate their eligibility for participation. Furthermore, there is a need to provide training programs for professionals to improve their skills and increase retrofit workforce capacity, which can play an essential role in addressing homeowners' concerns on technical problems.

Providing a bridge between homeowners and professionals can help in removing regulatory barriers and streamline the overall retrofitting process. For instance, establishing a central coordinator and consultant center can assist homeowners in evaluating retrofit opportunities, calculating retrofit costs, submitting rebate applications, verifying the retrofit quality, and conducting a post-retrofit survey. Meanwhile, developing a user-friendly tool (e.g., mobile app) can provide convenience for homeowners to explore retrofit packages and financial incentives most applicable to their homes.

In addition, the engagement of various stakeholders is also important for promoting energy retrofits. Therefore, it is recommended that social participation be encouraged by the inclusion of professionals and building owners in the working groups. Meanwhile, an organization, which functions as a mediator between the government and house owners, can coordinate various stakeholders and organize various activities to promote building renovation. In response to this situation, it is proposed that the participation of industry associations and non-profit organizations be increased in the energy retrofit process in BC, and that their roles be clearly defined, and for support to be given to them. With respect to working groups and industry associations, they should clearly define their relationships to avoid overlapping responsibilities.

| Advantages | Disadvantages |
|---|---|
| - It can help in increasing the industry capacity and implementing more retrofit projects at the same time.<br>- It can help in streamlining the retrofit process and improving work efficiency. | - It is time-consuming to be implemented.<br>- It cannot produce the direct benefit of embodied emission reductions.<br>- It requires extra financial supports from governments or utility providers. |

## 5.4 Establishment of an assessment and certification system for building embodied emissions

The establishment of an assessment and certification system is an effective way of dealing with the lack of building embodied emission data in energy retrofit projects. Furthermore, building performance certification (e.g., label, star, rating) systems can be used to improve the quality control of energy retrofit projects. Meanwhile, the data collected can be stored in a building material embodied emissions database for future usage. The assessment and certification system can provide real data support to regulation-based policies and financial support policies for building energy retrofits.

### 5.4.1 Life cycle carbon calculation and reporting

It is recommended that all retrofit projects calculate and report their life cycle carbon emissions using a standardized measure, separating embodied and operational carbon emissions. As such, stakeholders can understand the share of embodied emissions and operational emissions and which building types have the greatest carbon emission saving potentials, and thus develop optimal plans to reduce carbon emissions in retrofits. The retrofit industry should identify a standardized embodied emissions reporting tool (e.g., BEAM calculator for Part 9 buildings and Athena Impact Estimator for Part 3 buildings). At the same time, retrofit professionals should consider the fees associated with using the tool and collect input from industry stakeholders and develop guidelines for the selected tool to support the industry.

However, a lack of a comprehensive embodied emission database in BC poses a big challenge to calculate life cycle carbon emissions in retrofits. For example, the embodied emissions of many building insulation materials are still missing. While retrofit professionals can refer to other databases in the US and European countries, the embodied emission data may not be reliable due to different climate conditions, manufacturing methods, transportation systems, and landfill places.

### 5.4.2 Environmental Product Declaration (EPD)

Environmental product declaration (EPD) is a good example of reliable material information as EPD records data directly from manufacturers and companies, and it is developed strictly following ISO 21931 and EN 15643 at the building level. The overall goal of EPDs is to promote the supply of building products that are more environmentally friendly by communicating verifiable and accurate information, and to increase the potential of continuous environmental improvement. Based on ISO 14025, the detailed objectives of EPDs are as follows: 1) To provide LCA-based information and additional information on the environmental performance of products; 2) To help users and purchasers making informed comparisons between products; 3) To encourage improvement of environmental performance; 4) To provide information for evaluating the environmental impacts of products over their life cycle.

In particular, EPDs can provide embodied carbon emission data for building materials. A EPD materials database contains different construction materials environmental impact profiles for different products, technologies, suppliers and products. EPDs can be divided into three types based on system boundaries: cradle-to-gate, cradle-to-gate with options, and cradle-to-grave. A cradle-to-gate EPD only covers minimum information from the product stage. Cradle-to-gate with options EPD covers the information from the product stage, plus additional information from other stages. A cradle-to-grave type EPD covers all the life cycle stages as a minimum, and some benefits or loads beyond the system boundary might be included as well. As of May 2013, 556 PCR and 3614 EPD documents were published worldwide . Italy and Sweden have the highest number of products with EPDs, followed by Spain and Switzerland. It is recommended that BC industry develop EPDs for building construction and insulation materials and energy equipment commonly used in Canadian buildings. With a comprehensive EPD database, retrofit professionals can easily calculate life cycle carbon emissions of existing buildings and identify the buildings with highest emission reduction potential through energy retrofits.

| Advantages | Disadvantages |
| --- | --- |
| <ul><li>It can provide direct information associated with embodied emissions of various building materials and help retrofit planners compare different building materials and select the best one in in terms of the carbon and economic performance.</li></ul> | <ul><li>Establishing a comprehensive EPD database will be time-consuming and requires multiple building stakeholders to work together.</li><li>It requires extra efforts and time to complete life cycle carbon emission calculation and reporting.</li><li>It may increase the upfront cost of a retrofit project.</li></ul> |

## 5.5 Development of a comprehensive financial support system for retrofits

Financial incentives can effectively moderate stakeholders' concerns on economic issues such as high upfront cost, payback period, and return on investment. In this regard, it is necessary to provide financial support bundles for them. The bundles may comprise multiple forms of economic supports, such as subsidies, tax credits, and low-interest loans on conducting specific retrofit measures and rebates for purchasing energy-efficient equipment (e.g., heat pump). Furthermore, utility on-bill financing is also recommended, which can effectively address the concern of upfront cost. It is important to keep low administrative burden and interest cost and ensure that administrative and compliance systems can readily adapt to the new billing requirements. In order to assist homeowners in understanding the total actual costs of retrofit options, all available financial choices should be involved in retrofit costs shown to homeowners. While there are a few financial supports for energy retrofits from BC government and utility providers, such as Fortis BC and BC Hydro, further incentives need to be provided to reduce embodied emissions in the existing building sector.

### 5.5.1 Develop circular economy strategies

It is recommended that develop circular economy strategies to avoid embodied carbon by reducing resource extraction and repurposing materials already in circulation. In the building sector, circular economy strategies consist of salvaging and reusing building materials as well as re-using the core and shell of existing buildings. Developing a comprehensive circular economy strategy requires a huge shift from disposability to durability in the building sector, such as reducing the building area per occupant, prioritizing efficient occupancy of existing buildings, and removing provisions for private transportation [91].

### 5.5.2 Tax rebates for low carbon development

Governments can offer an annual property tax rebate for a set number of years to property owners who opt for low-carbon renovation rather than new construction. The amount of the rebate can be based on a quantification of the embodied carbon reduction, so that energy retrofit projects with greater relative embodied carbon reductions are eligible for larger rebates.

### 5.5.3 Carbon performance grants for retrofit projects

Municipalities can set special funds to award performance-conditioned grants for energy retrofit projects that achieve a clearly above market embodied carbon performance. Grants can be applied for during energy retrofit planning permission application, but they would be paid out only once project is completed, and performance achieved is possible to verify and audit.

### 5.5.4 Increase demolition permitting fees

Municipalities can increase demolition permitting fees for property owners applying to demolish buildings. The increases could be applied conditionally, according to building types, age, size, materials, carbon efficiency of the proposed replacement project, suitability of the building to deconstruction.

### 5.5.5 Incentives for manufacturers to reduce carbon

This strategy creates incentives for manufacturers in BC to reduce embodied carbon in their products. Possible pathways include:

- Property tax rebates for manufacturers in BC that demonstrate and quantify significant embodied carbon reductions in their main products.
- Property tax rebates for manufacturers in BC who meet a zero net carbon standard for the operation of their facilities, either by generating enough renewable, non-GHG emitting energy on-site to power their manufacturing processes, or procuring renewable, non-GHG emitting energy generated off-site.
- Direct grants for manufacturers for completing facility upgrades that significantly reduce carbon emissions, such as switching from a material or emissions intensive manufacturing process to a less material or emission-intensive alternative, or installing on-site renewable, non-GHG emitting energy generation.

### 5.5.6 Rebates for heavy-duty zero-emission vehicles

Governments can provide rebates for heavy-duty zero-emission vehicles to reduce embodied emissions in retrofits due to transportations. For example, the BC transportation department can provide rebates up to 100% of the price difference between an electric heavy-duty vehicle and a traditional gas-power heavy-duty vehicle. This strategy can encourage transport companies to purchase more zero-emissions vehicles and thereby reduce embodied emissions caused by the transportation of building materials and components.

### 5.5.7 Landfill tax on construction and demolition waste

It is recommended that government establishes a requirement for taxing all landfilled construction and demolition waste. Landfill tax will provide a broad financial incentive to avoid final disposal of all types of material streams. To have impact on construction and demolition waste, this tax should also be levied on aggregates.

### 5.6 Establishment of a comprehensive energy retrofit supply chain

A supply chain is a network between a company and its suppliers to produce and distribute a specific product or service. The entities in the supply chain include producers, vendors, warehouses, transportation companies, distribution centers, and retailers. Energy retrofits supply chain (ERSC) can be defined as a functional chain structure that includes the participants involved in each phase of the energy retrofit project life cycle. The participants in an energy retrofit project, such as the developer, contractor, materials manufacturer and supplier, are linked by cash, information and materials flow. In ERSC, the raw materials production and transportation phase may contribute the most carbon emissions. Regarding the carbon emissions of different ERSCs, the differences in the materials and component production, transportation and construction phases can be significant. Therefore, it is recommended that BC governments establish a comprehensive supply chain for implementing energy retrofits in BC, building relationships and facilitating the development of networks of manufacturers to support industrial symbiosis. A comprehensive supply chain can assist retrofit professionals in reducing unnecessary works, simplifying energy retrofit steps, optimizing building material manufacturing and transportation systems, and thereby decreasing embodied carbon emissions. At the same time, research and development (R&D) for the development of advanced ERSC should be supported by the governments or utility providers (e.g., Fortis BC and BC hydro) through special funds.

| Advantages | Disadvantages |
|---|---|
| - It can assist retrofit professionals in reducing unnecessary works, simplifying energy retrofit steps, optimizing building material manufacturing and transportation systems.<br>- It can have a significant impact on reducing embodied emissions in energy retrofits. | - It is complex and time-consuming to establish a supply chain since it requires multiple stakeholders work together, including academic researchers and industry people.<br>- It requires aside financial supports from the BC government and utility companies (e.g., Fortis BC and BC Hydro) |

## 6 Conclusion and recommendations

This research conducted a literature review and identification of BC policies and incentives for reduce overall GHG emissions from the existing building sector. Then, this research analyzed worldwide policies and incentives that reduce embodied carbon emissions in the existing building sector. After reviewing the two categories of retrofit policies, the author identified links and opportunities between existing BC strategies, tools, and incentives, and global embodied emission strategies. Finally, this research compiled key findings from all resources and provided policy recommendations for reducing embodied carbon emissions in retrofits in BC. The recommendations are as follows:

1. **Provide embodied carbon in retrofits related knowledge and information to the public:** Employ broadcast and internet and establish one-stop website to disseminate knowledge and information of embodied carbon emissions in energy retrofits. This can help in improving the public awareness of implementing energy retrofits to reduce carbon emissions in the existing building sector.

2. **Enhance building material emission performance:** Selecting salvaged and recycled materials, materials that sequester carbon, or materials that are manufactured and processed using low-carbon energy, and sourcing local supplies to avoid transportation emissions are effective and direct ways to reduce embodied carbon emissions in energy retrofits.

3. **Engage with energy retrofit professionals and organizations:** Engage with energy retrofit professionals and product manufacturers to learn what incentives would be most attractive to reduce embodied emissions in retrofits. Provide a bridge between homeowners and professionals can help in removing regulatory barriers and streamline the overall retrofitting process.

4. **Establish an assessment and certification system for building embodied emissions in retrofits:** Establishing an assessment and certification system is an effective way of dealing with the lack of building embodied emission data in energy retrofit projects. Furthermore, building performance certification (e.g., label, star, rating) systems can be used to improve the quality control of energy retrofit projects. Meanwhile, the data collected can be stored in a database associated with building material embodied emissions for future usage. A well-developed assessment and certification system can provide real data support for building energy retrofits.

5. **Develop a comprehensive financial support system for retrofits:** Explore financial incentives such as circular economy strategies, tax rebates for low carbon development, carbon performance grants, demolition permitting fees, incentives for manufacturers, rebates for heavy-duty zero-emission vehicles, and landfill tax on construction and demolition waste.

6. **Establish a comprehensive energy retrofit supply chain:** An energy retrofit supply chain is a functional chain structure that includes the participants involved in each phase of the energy retrofit project life cycle. Participants in an energy retrofit project, such as developers, contractors, material manufacturers and suppliers, are linked by cash, information and materials flow. A comprehensive supply chain can assist retrofit professionals in reducing unnecessary works, simplifying energy retrofit steps, optimizing building material manufacturing and transportation systems, and thereby decreasing embodied carbon emissions in retrofits.

485–496. https://doi.org/10.1016/j.enbuild.2011.12.028.

[44] X. Kong, S. Lu, Y. Wu, A review of building energy efficiency in China during "Eleventh Five-Year Plan" period, Energy Policy. 41 (2012) 624–635. https://doi.org/10.1016/j.enpol.2011.11.024.

[45] H. Tasdoven, B.A. Fiedler, V. Garayev, Improving electricity efficiency in Turkey by addressing illegal electricity consumption: A governance approach, Energy Policy. 43 (2012) 226–234. https://doi.org/10.1016/j.enpol.2011.12.059.

[46] S. Boyle, DSM progress and lessons in the global context, Energy Policy. 24 (1996) 345–359. https://doi.org/10.1016/0301-4215(95)00142-5.

[47] EconStor: Cross-country econometric study on the impact of fiscal incentives on FDI, (n.d.).

[48] K. Kempa, U. Moslener, Climate policy with the chequebook - An economic analysis of climate investment support, Econ. Energy Environ. Policy. 6 (2017) 111–129. https://doi.org/10.5547/2160-5890.6.1.kkem.

[49] E.S. Kirschen, Economic policy in our time, (1964).

[50] A. Rana, R. Sadiq, M.S. Alam, H. Karunathilake, K. Hewage, Evaluation of financial incentives for green buildings in Canadian landscape, Renew. Sustain. Energy Rev. 135 (2021) 110199. https://doi.org/10.1016/j.rser.2020.110199.

[51] P. Bonifaci, S. Copiello, Incentive Policies for Residential Buildings Energy Retrofit: An Analysis of Tax Rebate Programs in Italy, in: A. Bisello, D. Vettorato, P. Laconte, S. Costa (Eds.), Smart Sustain. Plan. Cities Reg. Sspcr 2017, 2018: pp. 267–279. https://doi.org/10.1007/978-3-319-75774-2_19.

[52] E. Baldoni, S. Coderoni, M. D'Orazio, E. Di Giuseppe, R. Esposti, The role of economic and policy variables in energy-efficient retrofitting assessment. A stochastic Life Cycle Costing methodology, Energy Policy. 129 (2019) 1207–1219. https://doi.org/10.1016/j.enpol.2019.03.018.

[53] B.A. Brotman, The impact of corporate tax policy on sustainable retrofits, J. Corp. Real Estate. 19 (2017) 53–63. https://doi.org/10.1108/jcre-02-2016-0011.

[54] X. Liang, T. Yu, J.K. Hong, G.Q. Shen, Making incentive policies more effective: An agent-based model for energy-efficiency retrofit in China, Energy Policy. 126 (2019) 177–189. https://doi.org/10.1016/j.enpol.2018.11.029.

[55] Y. Tan, T. Luo, X. Xue, G.Q. Shen, G. Zhang, L. Hou, An empirical study of green retrofit technologies and policies for aged residential buildings in Hong Kong, J. Build. Eng. 39 (2021). https://doi.org/10.1016/j.jobe.2021.102271.

[56] Roadmap 2030 | CleanBC, (n.d.). https://cleanbc.gov.bc.ca/ (accessed May 24, 2022).

[57] Retrofitting, (n.d.). https://www.nrcan.gc.ca/energy-efficiency/buildings/existing-buildings/retrofitting/20707 (accessed May 21, 2022).
23

# Appendix

List of BC policies and incentives

| Direction and Command | Assessment and Disclosure | Research and Service | Financial incentives |
|---|---|---|---|
| Minor, Major and Deep Retrofit Suggestions for Homes https://www.nrcan.gc.ca/energy-efficiency/buildings/existing-buildings/retrofitting/20707 | EnerGuide Rating System https://www.bchydro.com/powersmart/business/programs/new-home/energuide-for-new-houses.html | EfficiencyBC program https://www.toolkit.bc.ca/Program/EfficiencyBC | NRCan Grants for Home Retrofits https://www.nrcan.gc.ca/energy-efficiency/homes/canada-greener-homes-grant/start-your-energy-efficient-retrofits/plan-document-and-complete-your-home-retrofits/eligible-grants-for-my-home-retrofit/23504 |
| Clean BC Roadmap to 2030 https://cleanbc.gov.bc.ca/ | | CleanBC Better Homes https://betterhomesbc.ca | CleanBC Better Buildings rebates https://betterbuildingsbc.ca |
| | | CleanBC Better Buildings https://betterbuildingsbc.ca | CleanBC Better Homes rebates https://betterhomesbc.ca |
| | | CleanBC Commercial Express Program https://betterbuildingsbc.ca/incentives/cleanbc-commercial-express-program | Clean BC Better Homes Low-interest Financing program https://betterhomesbc.ca/rebates/financing |
| | | | Fortis BC retrofit rebates https://www.fortisbc.com/rebates-and-energy-savings/rebates-and-offers?l= |
| | | | BC Hydro retrofit rebates https://www.bchydro.com/powersmart/residential/rebates-programs/home-renovation.html |
| | | | BC Housing Energy Efficiency Retrofit Program |

List of global Policies that reduce Embodied Emissions adapted to existing buildings

| Building Regulations | Procurement | Waste & Circularity | Financial Policies |
|---|---|---|---|
| Construction materials efficiency declaration | Carbon limits for building materials procurement | Design for disassembly and adaptability criteria | Tax rebates for low carbon development |
| Expedited permitting for low carbon project | Requirement of recycled aggregates | Mandatory pre-demolition audits and data sharing | Link land use fees to project life cycle carbon |
| Prohibiting extremely high emitting materials | Require use of certified wood products | Mandatory material takeback program | Carbon performance grants for projects |
| Life cycle carbon calculation and reporting | Circular materials purchasing strategy | Soil coordination for mass storage and reuse | Include embodied carbon in climate action plan |
| | | Information on adaptability and waste reduction | Increase demolition permitting fees |
| | | Materials longevity policy | Incentives or manufacturers to reduce carbon |
| | | | Landfill tax on construction and demolition waste |